
%
\hyphenation{over-ion-isation}

\def\ee#1{\cdot 10^{#1}}


\def\feh{{\rm [Fe/H]}}
\def\teff{$T_{\rm eff}$}
\def\bi{B\,{\sc i}}
\def\bii{B\,{\sc ii}}

%
%
  \MAINTITLE={ A NLTE study of neutral boron in solar-type stars }
%
%
%
  \SUBTITLE={ ????? }       
  \AUTHOR={ Dan Kiselman@1@2 }         
<, Nextauthor@number>
%
  \INSTITUTE={ @1 Astronomiska observatoriet, Box 515, S-751~20~~Uppsala,
Sweden
@2 Nordita, Blegdamsvej 17, DK-2100 K\o benhavn~\O, Denmark}
%
  \ABSTRACT={ The formation of the resonance lines of neutral boron
in solar-type stellar atmospheres is investigated taking into
account effects of departures from local thermodynamic equilibrium
(NLTE effects). The latter are due to a combination of overionisation
and optical pumping in resonance lines, both caused
by the hot, non-local, ultraviolet radiation fields in the
line-forming regions. They lead to an underestimation of the boron abundance
when analysis methods assuming local thermodynamic
equilibrium (LTE) are used. The abundance
correction, for the 249.7~nm resonance line,
amounts to $+$0.6 dex for the metal-poor star HD140283 and $+$0.4 dex for
Procyon.
No significant NLTE effects are predicted for the Sun.
Applying the abundance correction on the results for HD140283
of Duncan et al. (1992) leads to a B/Be ratio well above the minimum
value required by spallation production of beryllium.
The reliability of the results in view of atomic and atmospheric uncertainties
is discussed. With the possible exception of photospheric inhomogeneity,
it seems unlikely that these could remove the effect for HD140283.
 }       
  \KEYWORDS={ Line: formation -- Sun: abundances -- Stars: abundances --
Stars: atmospheres -- Galaxy: abundances --
Boron }       
  \THESAURUS={ 02.12.1, 06.01.1, 08.01.1, 08.01.3, 10.01.1 }
  \OFFPRINTS={ ????? }      
  \DATE={ ????? }           
%
\maketitle
\titlea{Introduction}

This paper discusses the statistical equilibrium of neutral boron
for three model atmospheres representing stars for which neutral boron
 lines have been used
to determine abundances. These determinations have all been made using the
assumption of local thermodynamic equilibrium (LTE). The ambition
here is to investigate whether departures from LTE -- NLTE effects --
exist for neutral boron in solar-like atmospheres, and whether these
effects are large enough to have a significant impact on the derived
abundances.
Of the three stars to be investigated, the emphasis will be on the extreme halo
star
HD140283. The other two stellar models represent the F dwarf Procyon and the
Sun.
The current paper supersedes earlier reports on this subject (Kiselman 1992,
1993a)

\titleb{Background}
The boron abundance of HD140283 was determined by
Duncan et al. (1992, hereafter DLL92), from Hubble Space Telescope spectra.
Considerable interest has been raised concerning the
ratio of boron to beryllium in such metal-poor stars.
Gilmore et al. (1991) found an unexpectedly high
beryllium abundance in HD140283, and it has been discussed whether
this may represent a primordial abundance. If so, the standard
model for cosmological nucleosynthesis would have to be changed
(see Pagel 1991). There is,
however, the possibility that the beryllium in HD140283 instead was
produced by interstellar spallation processes in the early Galaxy.
Spallation would also produce appreciable amounts of boron,
thus the boron abundance of the star can be used to settle the question.
DLL92 found that spallation production should give
an abundance ratio ${\rm B/Be} >10$. Combining their (LTE) boron abundance
result with the beryllium abundance of Gilmore et al. (1991), they arrive
at ${\rm B/Be} = 10$, which is barely consistent with spallation production.
This ratio may be an upper limit since DLL92 do not
rule out the possibility that the boron line is blended with an unidentified
line.

The Sun's boron abundance was determined by Kohl et al. (1977) using
rocket spectra of the boron resonance line at 249.7~nm. Using
the same spectral line, Lemke et al. (1993, hereafter LLE93) derived the
boron abundance of Procyon from Hubble Space Telescope spectra.
They found a boron depletion by a factor
of at least 3 compared to the Sun, which is much less than the
corresponding depletion
of Li and Be in Procyon. LLE93 propose that the most probable explanation
for this depletion in Procyon is envelope mixing by rotation-induced currents
to levels where boron is destroyed only very slowly.

\titleb{The boron atom and expected NLTE effects}
With the NLTE-calculation tools available at present, it is not possible
to carry out the kind of
synthetic spectroscopy that is necessary for a detailed spectral analysis in
crowded spectral regions. My ambition here is not, however,
 to perform a detailed spectral analysis of the three stars, but rather to
investigate the possible NLTE effects and to suggest correction factors to the
LTE results.

Neutral boron has an ionisation potential of 8.3~eV; the ionisation degree
of boron in solar-type stellar photospheres thus varies from largely
neutral in the Sun to almost completely ionised in hotter or more metal-poor
atmospheres, cf. Fig. 3.

The excited levels have energies from 3.5~eV and upwards (cf. Fig. 1).
So far, only the resonance doublet at 250~nm has
 been employed for abundance analysis of solar-type
stars, but Johansson et al.  (1993) point out that another
resonance doublet, at 209~nm, may be more useful, especially
when it comes to determining isotope ratios.
These two resonance multiplets will be analysed here.
For each of them results will be given
for the component that is least
affected by blending, viz. the lines at (air) wavelengths 249.68~nm and
208.96~nm.

Inspired by the review of Rutten (1990) we
can try to predict what type of NLTE effects
that may be important for neutral boron and its resonance lines.
The radiation field in solar-type photospheres tends to be
hotter than the Planckian value (as given by the local electron temperature)
in the blue and near UV. This can lead to overionisation if there
are enough populated levels with ionisation potentials
around 2-5~eV, a range where we find the first excited levels of neutral boron.
Such an overionisation naturally leads to a weakening of neutral absorption
lines via a lowering of the line opacity.
But all ionisations must be counterbalanced by recombinations and it
is the competition between these processes that sets the ionisation balance.
For the
alkali atoms in the solar case, Bruls et al. (1992) showed that
 overrecombination to the
higher levels overcomes the overionisation from the lower, resulting
in a total underionisation relative to LTE. The overrecombination flow in this
case is driven by photon losses in the cascade of infrared transitions
 -- a process called "photon suction".
Another expected effect is optical pumping in the \bi\ resonance lines
 by the
hot non-local radiation field in the ultraviolet,
causing the upper levels of the lines to be overpopulated relative to
the Boltzmann
excitation equilibrium and thus a line source function $S_l > B$, leading
to a weakening of the absorption lines relative to the LTE case.
This could happen at low abundances when the lines are weak.
 When strong enough, the lines
influence their own radiation field so that $S_l$ is lowered.
It is the interplay between these processes -- overionisation,
photon suction in
the lines between the excited levels, and optical
pumping in the resonance lines -- that can be expected to cause departures
from LTE in the case of neutral boron.

\titlea{Techniques}
This section presents the techniques used for computing the model atmospheres
and for the NLTE line calculations.

\titleb{Definitions}
Boron abundances are expressed on the traditional logarithmic scale
relative to hydrogen: $A_{\rm B} = \lg N_{\rm B}/N_{\rm H} + 12$.
(Logarithms to base ten are written as $\lg$, natural logarithms
as $\ln$.)
The general metal content of the stars is expressed relative
to solar by the common bracket notation: $\feh =
\lg (N_{\rm Fe}/N_{\rm H}) - \lg (N_{\rm Fe}/N_{\rm H})_\odot$.
Equivalent widths are given in pm (1~m\AA = 0.1~pm).
When referencing and displaying results as a function of depth in the model
atmospheres, the continuum optical depth at 500~nm, $\tau_5$, will be used.

\titleb{Model atmospheres}
The theoretical homogeneous atmospheric models used here were computed with
the Edvardsson et al.  (1993) version of the
MARCS code (Gustafsson et al.  1975). This program computes line opacities
in the blue and ultraviolet part of the spectrum using opacity sampling,
taking into account millions of spectral lines from the recent
data produced by R. Kurucz (e.g. Kurucz 1992). The models yield surface fluxes
which show good agreement with observations of solar-type stars
(Edvardsson et al. 1993). They are computed assuming LTE and have
 no chromospheres.
The computed mean intensities for each wavelength and depth point
were saved to be used
for the calculation of photoionisation rates.

The atmospheric parameters (Table 1) were chosen to be similar to those used
in the LTE analyses of DLL92, LLE93, and Kohl et al. (1977).

\begfig 6.5 cm
\figure{1}{ A Grotrian diagram of the neutral boron model
atom used in this paper showing energy levels and
radiative transitions that are treated in detail. The resonance lines
of interest for abundance analysis are labeled with their wavelengths
in nm. Data for the levels are given in Table 2 according to their numbers
}
\endfig

\begfig 6.5 cm
\figure{2}{ The departure of the radiation field from the local Planck
function as
calculated by the model atmosphere program. The smoothed data, which
was used for computing photoionisation rates, is
plotted for representative line-forming depths in the three
stellar models: $\lg \tau_5 = -2.6$ for the Sun,
 $\lg \tau_5 = -1.2$ for Procyon and HD140283.
The positions of the ionisation edges for the ground state and
first excited states of neutral boron are indicated. The edge belonging
to level 2 is situated shortwards to that of the ground state since
this level ionises to an excited state of \bii
}
\endfig

\titleb{NLTE calculation techniques}
For the NLTE calculations, version 2.0 of the code MULTI (Carlsson 1986)
was used. The original version of this code uses the approximative
lambda operator scheme
developed by Scharmer (1981) and Scharmer \& Carlsson (1985) to
solve the coupled equations of statistical equilibrium and radiative
transfer for a multi-level atom which is treated as a trace element
in a given atmosphere. The current version also features a lambda operator
of a more local nature (Carlsson 1991). This operator was used
when computing the results presented here, a choice motivated
by its smaller demand for computer memory. (The final converged
solutions do not depend on the kind of operator used.)

The code was used with a more elaborate package for computing continuous
background opacities than what was available in earlier versions of the
program.
These are essentially the same continuous opacities
as in the MARCS program (Gustafsson et al. 1975).
It should be noted that this opacity package only contains continuous
opacities -- no opacity sampling as in the atmosphere code --
 and that the opacities are computed assuming LTE. This
restriction is of importance for the current study. Without line
opacities, the treatment of photoionisations,
for which the blue and ultraviolet radiation fields are important, can
be seriously in error. Therefore, use was made of the MULTI
feature which permits the treatment of radiative transitions as
transitions at fixed rates which are calculated from a specified
radiation field. These radiation fields, in the form of
$\ln (J_\nu/B_\nu)$ values specified for all depth points and a large
range of wavelength points, were produced by the model atmosphere
code, which includes the ultraviolet line blanketing via opacity sampling.
Before being used as input
to MULTI, the values were smoothed in frequency
in the same way as the photoionisation cross sections (see Fig. 2 and
below).

For the detailed calculations of all the ultraviolet resonance lines,
 the problem
of background opacities is more difficult since we
cannot then rely on statistical methods. A proper treatment requires
that a detailed line list in the relevant
spectral region is incorporated with the background opacities.
Some tests
will be presented here where the line-blanketing effects have been treated
in the same way as for the photoionisation calculations.

The LTE equivalent widths that are given here were also calculated
in the MULTI runs. The program sets the level population to the
appropriate Boltzmann and Saha distributions and performs the formal
line calculations. LTE quantities are marked with $^*$ through
this paper. When departure coefficients for level populations
are used, they are defined as $b_i = n/n^*$.
Due to the nature of the effects discussed in this paper
it is sometimes more illustrative to use a population depletion coefficient
which is defined as $d_i = 1/b_i = n^*/n$. In the same way,
the NLTE effect on the equivalent width of a line will sometimes
be given as $W^*/W$ instead of the more usual $W/W^*$.

\begtabfull
\tabcap{1}{Stellar parameters. [Fe/H] is expressed relative
to a solar Fe abundance of 7.51}
\halign{#\hfil\quad&#\hfill\quad\ &#\hfill\quad\ &\hfill#\quad\
&#\hfill\quad\ \cr
\noalign{\hrule\medskip}
Star & \teff  & $\lg g$~  & \feh & $\xi$~~~~  \cr
     & [K]    & [m\,s$^{-2}$]&  & [km\,s$^{-1}$]\cr
\noalign{\medskip\hrule\medskip}
Sun  & 5780      & 2.44                & ~~0.00   & 1.15      \cr
Procyon & 6700   & 2.03                & $-$0.04  & 2.5     \cr
HD140283 & 5640  & 1.30                & $-$2.63  & 1.5      \cr
\noalign{\medskip\hrule}
}
\endtab
\begtabfull
\tabcap{2}{ Energy levels included in the model atom. The numbers $i$ are
used for reference in the text and figures}
\halign{#\hfil\quad&#\hfill\quad\ &#\hfill\quad\ &#\hfill\ &#\ &#\ &#\ \cr
$i$   & $E$       & $g_i$  &   \cr
      & [eV]      &        &               \cr
\noalign{\hrule\medskip}
   0  &  0.000000 &  2     &  \bi\ \ $2s^2\ 2p$& $^2P^O_{1/2}$    \cr
   1  &  0.001895 &  4     &  \bi\ \ $2s^2\ 2p$& $^2P^O_{3/2}$    \cr
   2  &  3.572349 & 12     &  \bi\ \ $2s\ 2p^2$&     $^4P^E         $  \cr
   3  &  4.964326 &  2     &  \bi\ \ $2s^2\ 3s$& $^2S^E     $\cr
   4  &  5.933567 & 10     &  \bi\ \ $2s\ 2p^2$& $^2D^E     $   \cr
   5  &  6.027368 &  6     &  \bi\ \ $2s^2\ 3p$& $^2P^O     $   \cr
   6  &  6.790335 & 10     &  \bi\ \ $2s^2\ 3d$& $^2D^E     $   \cr
   7  &  6.820371 &  2     &  \bi\ \ $2s^2\ 4s$& $^2S^E     $\cr
   8  &  7.164709 &  6     &  \bi\ \ $2s^2\ 4p$& $^2P^O     $   \cr
   9  &  7.438216 & 10     &  \bi\ \ $2s^2\ 4d$& $^2D^E     $   \cr
  10  &  7.442959 & 14     &  \bi\ \ $2s^2\ 4f$& $^2F^O     $   \cr
  11  &  7.457261 &  2     &  \bi\ \ $2s^2\ 5s$& $^2S^E     $\cr
  12  &  7.516480 &  6     &  \bi\ \ $2s^2\ 5p$& $^2P^O     $   \cr
  13  &  7.699234 &  2     &  \bi\ \ $2s^2\ 6s$& $^2S^E     $\cr
  14  &  7.742071 &  6     &  \bi\ \ $2s^2\ 6p$& $^2P^O     $   \cr
  15  &  7.747212 & 10     &  \bi\ \ $2s^2\ 5d$& $^2D^E     $   \cr
  16  &  7.751125 & 14     &  \bi\ \ $2s^2\ 5f$& $^2F^O     $   \cr
  17  &  7.871326 &  6     &  \bi\ \ $2s^2\ 7p$& $^2P^O     $   \cr
  18  &  7.880579 &  2     &  \bi\ \ $2s\ 2p^2$& $^2S^E    $\cr
  19  &  7.915799 & 10     &  \bi\ \ $2s^2\ 6d$& $^2D^E     $   \cr
  20  &  7.918481 & 14     &  \bi\ \ $2s^2\ 6f$& $^2F^O     $   \cr
  21  &  7.952164 &  6     &  \bi\ \ $2s^2\ 8p$& $^2P^O     $   \cr
  22  &  7.954395 &  2     &  \bi\ \ $2s^2\ 7s$& $^2S^E     $\cr
  23  &  8.006222 &  6     &  \bi\ \ $2s^2\ 9p$& $^2P^O     $   \cr
  24  &  8.017246 & 10     &  \bi\ \ $2s^2\ 7d$& $^2D^E     $   \cr
  25  &  8.019332 & 14     &  \bi\ \ $2s^2\ 7f$& $^2F^O     $   \cr
  26  &  8.033264 &  2     &  \bi\ \ $2s^2\ 8s$& $^2S^E     $\cr
  27  &  8.044050 &  6     &  \bi\ \ $2s^2\ 10p$& $^2P^O    $   \cr
  28  &  8.083217 & 10     &  \bi\ \ $2s^2\ 8d$& $^2D^E     $   \cr
  29  &  8.084766 & 14     &  \bi\ \ $2s^2\ 8f$& $^2F^O     $   \cr
  30  &  8.298095 &  1     &  \bii\  $2s^2$    & $^1S^E     $ \cr
\noalign{\medskip\hrule}
}
\endtab

\begtabfull
\tabcap{3}{Parameters of the \bi\ resonance lines that are of observational
interest. The lines marked with * are the ones discussed in the
paper, since they are least affected by blending }
\halign{#\hfil\quad&#\hfil\quad&#\hfill\quad\ &#\hfill\quad\ &#\hfill\quad\
&#\hfill\quad\ \cr
lower & upper & $\lambda_{\rm air}$ & $f_{ij}$ & $g_i$ \cr
$i$ & $j$ &    [nm]             &          &       \cr
\noalign{\medskip\hrule\medskip}
  0 & 4   &   208.89              & 0.0475   &  2    \cr
  1 & 4   &   208.96*              & 0.0450   &  4    \cr
  0 & 3   &   249.68*              & 0.0830   &  2    \cr
  1 & 3   &   249.77              & 0.0830   &  4    \cr
\noalign{\medskip\hrule}
}
\endtab

\begfig 14 cm
\figure{3} { The fraction of ionised boron in the three stellar model
atmospheres. The lines show the NLTE results of this paper,
the stars are LTE values
}
\endfig

\titlea{Model atom}

The neutral boron model atom was compiled from laboratory data and
from theoretical data computed within the Opacity
Project (hereafter denoted OP, see Seaton 1987; Berrington et al. 1987).
 The OP data for the photoionisation cross sections
and the oscillator strengths was extracted from the TOPbase (Cunto et al.
1993).

\titleb{Atomic levels}
Energy levels were taken from the laboratory data compilation of
Bashkin \& Stoner (1978),
with the exception of the higher among the $^2$P levels for which
the values were taken from the OP data.
The continuum was represented by the ground state of \bii.
Data for the levels are given in Table 2. The levels
will be referred to with their numbers in that table.

\titleb{Photoionisation cross sections}
The photoionisation cross sections were all OP data.
Data for the $^4$P level was not included in the TOPbase, but was kindly
made available by Dr.~K.~Berrington. The $^2$F levels' cross sections
were copied from the levels of $^2$P with similar energies.

In the current work, the cross-sections are multiplied
with mean intensity values and integrated over frequency. Since
the mean intensities come from a model atmosphere program with
opacity sampling, it is important that the mean intensities
and the cross sections have comparable frequency distributions to
avoid spurious sampling effects.
Therefore, the cross-sections were smoothed to a frequency grid of
about $5\ee{13}$~Hz spacing. The smoothing was done
via trapezoidal integration over the appropriate intervals.

\begfigwid 11 cm
\figure{4}{ Upper panels show departure coefficients $b_i = n/n^*$ for
the ground state and the first six excited levels of neutral boron.
Lines with crosses
show the depletion factor for the neutral atoms. The horizontal bar
has ticks at total optical depth 0.1 and 1.0 in the centre of the
249.7~nm line.\hfill\break
Lower panels show deviations of line source functions $S_l$ from
local Planckian values $B$ (solid lines). In this case the relation
$S_l/B = b_u/b_l$ holds. Open symbols denotes
the $J_C/B$ ratio for the different lines, where $J_C$ is the mean
intensity in the continuum.
The 249.7~nm line connects levels 0 and 3, 209.0~nm connects
1 and 4 ($b_1=b_0$), 182.6~nm connects 0 and 6 \hfill\break
}
\endfig

\titleb{Oscillator strengths}
For the resonance lines studied here, precision laboratory
data were available from Johansson et al. (1993), see Table 3.
Oscillator strengths for the rest of the radiative transitions
were taken from OP data.
Lines with $f < 0.001$ and
lines of extremely long wavelength were excluded.

\titleb{Line broadening}
Radiation damping constants were calculated from the lifetimes of the levels.
The Uns\"old (1955, p. 333)
 prescription was used to treat van der Waals damping.

\titleb{Treatment of fine structure}
The ground state was split in the two fine structure levels.
These were coupled strongly by setting the collisional rates
to very large values. The rest of the levels were treated as terms,
and the radiative transitions connecting them as merged multiplets.
In the splitting of the OP data for the resonance lines into doublets,
it was assumed that both components of each doublet had the same
f-value.

Hyperfine splitting and isotopic splitting were neglected.

\titleb{Collisions with electrons}
Detailed theoretical data for electronic b-b cross sections
has been calculated by Nakazaki \& Berrington (1991) for the transitions
between the ground state (0, 1) and the four lowest excited states (2, 3, 4,
5),
as well as between the $^4$P level (2) and the two next higher levels (3, 4).
The appropriate rate coefficients were
calculated from the cross sections as given in tables and plots in that paper.
In a few cases some rough extrapolation had to be employed
for the high energy tails of the cross sections.
For the transitions between the ground state (0, 1) and (6, 7, 8),
Dr.~K.~Berrington kindly provided rate coefficients which were not given
in the paper. Nakazaki \& Berrington
(1991) compare their data with published laboratory measurements and
note good agreement in shape between the cross section -- energy
relations if the laboratory data is divided with a factor
between 4 and 6. They give reasons why the laboratory data could
have such systematic offsets.
The theoretical values are employed here
 as published. For the remaining transitions, the rates
were calculated using the approximation of Van Regemorter (1962)
with the appropriate $f$-values when available. For other (forbidden)
transitions $f = 0.01$ was used.

Collisional ionisation rates by electrons were calculated from
the hydrogenic approximations given by Allen (1973).

\titleb{Collisions with neutral atoms}
The importance of inelastic collisions with neutral atoms,
notably hydrogen,
has been a subject of concern in stellar studies for some time. There seems to
be virtually no experimental or detailed theoretical data
in the relevant energy ranges. I have tested the analytic
expressions for the ionisation cross sections of Kunc \& Soon (1991).
The formulae seem to give a reasonable agreement with
existing laboratory data for the cases demonstrated
by these authors, but the examples are for much higher
energy than what is of interest here.
The appropriate rate coefficients were calculated and are
included in the calculations presented here. However, these
collisional rates turn out to be several orders of magnitude smaller
than the corresponding rates due to electron collisions.

Regarding collisionally induced bound-bound transitions, there are
reasons to believe that the rates given by Drawin (1969) (which
in astrophysical applications are used in the form presented
by Steenbock \& Holweger (1984)) are overestimated (Caccin et al.
1993; Lambert 1993; Kiselman 1993b). Such processes are not
included here, with the exception of one test to be described later.

\titleb{Final model}
The final model atom, whose term diagram appears as Fig.~1,
 includes 30 bound
levels and one continuum level, 114 radiative bound-bound radiative
transitions treated in detail and 30 photoionisation
continua which are treated as fixed rates. Subsequent experiments
later showed that it would be possible to use
a reduced number of levels while preserving
the essential properties of the atom for the line calculations.
Nevertheless, the full model was employed in all calculations
presented here.
Note that most of the oscillator strengths, the photoionisation
cross sections and the electron collisional cross sections for
transitions between the lowest levels are of substantial accuracy
 and should not
cause any errors greater than what is expected from atmospheric
uncertainties, etc. The possible trouble spots
 of the model atom will be discussed later, when
the most important of the quantities for the current problem have
been identified.

\begtabfull
\tabcap{4}{The NLTE effects on equivalent widths and derived abundances,
HD140283. $A_{\rm B}$(LTE) is the result of an LTE analysis when the
true abundance is $A_{\rm B}$ and the observed equivalent
width is $W$. $W^*$ is the expected equivalent width for abundance
$A_{\rm B}$ if the
lines were formed in LTE. $\Delta A_{\rm B} = A_{\rm B} - A_{\rm B}$(LTE): "the
NLTE correction to the abundance"
} \halign{#\hfil&\hfill#\ &\hfill#\ &\hfill#\ &\hfill#\ &\hfill#\
&\hfill#\ \cr \noalign{\hrule\medskip}
         & $A_{\rm B}^{\rm LTE}$ & $A_{\rm B}$  & $W$ [pm] & $W/W^*$
& $W^*/W$ & $\Delta A_{\rm B}$  \cr
\noalign{\medskip\hrule\medskip}
209.0~nm  &  $-$1.50 &  $-$0.86 &     0.03 &    0.24 &   4.23 &   $+$0.64 \cr
          &  $-$1.00 &  $-$0.36 &     0.09 &    0.24 &   4.16 &   $+$0.64 \cr
          &  $-$0.50 &     0.14 &     0.27 &    0.26 &   3.87 &   $+$0.64 \cr
          &     0.00 &     0.64 &     0.81 &    0.31 &   3.22 &   $+$0.64 \cr
          &     0.50 &     1.16 &     2.12 &    0.43 &   2.30 &   $+$0.66 \cr
          &     1.00 &     1.67 &     4.38 &    0.62 &   1.60 &   $+$0.67 \cr
          &     1.50 &     2.06 &     6.55 &    0.77 &   1.29 &   $+$0.56 \cr
\noalign{\medskip\hrule\medskip}
249.7~nm  &  $-$1.50 &  $-$0.94 &     0.03 &    0.28 &   3.61 &   $+$0.56 \cr
          &  $-$1.00 &  $-$0.44 &     0.11 &    0.28 &   3.54 &   $+$0.56 \cr
          &  $-$0.50 &     0.06 &     0.33 &    0.30 &   3.34 &   $+$0.56 \cr
          &     0.00 &     0.56 &     0.98 &    0.35 &   2.86 &   $+$0.56 \cr
          &     0.50 &     1.07 &     2.56 &    0.47 &   2.13 &   $+$0.57 \cr
          &     1.00 &     1.56 &     5.23 &    0.66 &   1.52 &   $+$0.56 \cr
          &     1.50 &     2.00 &     7.74 &    0.80 &   1.25 &   $+$0.50 \cr
\noalign{\medskip\hrule}
}
\endtab

\begtabfull
\tabcap{5}{The NLTE effects on equivalent widths and derived abundances,
Procyon} \halign{#\hfil&\hfill#\ &\hfill#\ &\hfill#\ &\hfill#\ &\hfill#\
&\hfill#\ \cr
\noalign{\hrule\medskip}
         & $A_{\rm B}^{\rm LTE}$ & $A_{\rm B}$ & $W$ [pm] & $W/W^*$
& $W^*/W$   & $\Delta A_{\rm B}$
\cr \noalign{\medskip\hrule\medskip}
209.0~nm &   0.50 &   0.94 &     0.14 &    0.38 &   2.66 &   $+$0.44 \cr
         &   1.00 &   1.44 &     0.42 &    0.39 &   2.55 &   $+$0.44 \cr
         &   1.50 &   1.94 &     1.23 &    0.44 &   2.26 &   $+$0.44 \cr
         &   2.00 &   2.44 &     3.14 &    0.56 &   1.77 &   $+$0.44 \cr
         &   2.50 &   2.93 &     6.05 &    0.75 &   1.34 &   $+$0.43 \cr
         &   3.00 &   3.35 &     8.55 &    0.87 &   1.15 &   $+$0.35 \cr
         &   3.50 &   3.71 &    10.48 &    0.94 &   1.07 &   $+$0.21 \cr
 \noalign{\medskip\hrule\medskip}
249.7~nm &   0.50 &   0.90 &     0.14 &    0.40 &   2.48 &   $+$0.40 \cr
         &   1.00 &   1.40 &     0.43 &    0.42 &   2.38 &   $+$0.40 \cr
         &   1.50 &   1.90 &     1.26 &    0.47 &   2.12 &   $+$0.40 \cr
         &   2.00 &   2.39 &     3.25 &    0.60 &   1.68 &   $+$0.39 \cr
         &   2.50 &   2.87 &     6.33 &    0.76 &   1.31 &   $+$0.37 \cr
         &   3.00 &   3.30 &     9.14 &    0.88 &   1.14 &   $+$0.30 \cr
         &   3.50 &   3.69 &    11.42 &    0.94 &   1.07 &   $+$0.19 \cr
 \noalign{\medskip\hrule\medskip}
}
\endtab

\begtabfull
\tabcap{6}{The NLTE effects on equivalent widths and derived abundances, the
Sun}
\halign{#\hfil&\hfill#\ &\hfill#\ &\hfill#\ &\hfill#\ &\hfill#\ &\hfill#\ \cr
\noalign{\hrule\medskip}
         & $A_{\rm B}^{\rm LTE}$ & $A_{\rm B}$ & $W$ [pm] & $W/W^*$  & $W^*/W$
& $\Delta A_{\rm B}$  \cr
\noalign{\medskip\hrule\medskip}
209.0~nm   &   0.50 &   0.73 &     0.52 &    0.63 &   1.60 &   $+$0.23 \cr
           &   1.00 &   1.24 &     1.47 &    0.67 &   1.49 &   $+$0.24 \cr
           &   1.50 &   1.77 &     3.40 &    0.76 &   1.32 &   $+$0.27 \cr
           &   2.00 &   2.26 &     5.71 &    0.87 &   1.15 &   $+$0.26 \cr
           &   2.50 &   2.66 &     7.52 &    0.94 &   1.07 &   $+$0.16 \cr
           &   3.00 &   3.07 &     9.18 &    0.97 &   1.03 &   $+$0.07 \cr
           &   3.50 &   3.52 &    11.17 &    0.99 &   1.01 &   $+$0.02 \cr
\noalign{\medskip\hrule\medskip}
249.7~nm  &   0.50 &   0.66 &     0.35 &    0.71 &   1.41 &   $+$0.16 \cr
          &   1.00 &   1.16 &     1.01 &    0.75 &   1.34 &   $+$0.16 \cr
          &   1.50 &   1.66 &     2.46 &    0.83 &   1.21 &   $+$0.16 \cr
          &   2.00 &   2.12 &     4.50 &    0.92 &   1.09 &   $+$0.12 \cr
          &   2.50 &   2.53 &     6.43 &    0.98 &   1.02 &   $+$0.03 \cr
          &   3.00 &   2.97 &     8.25 &    1.01 &   0.99 &   $-$0.03 \cr
          &   3.50 &   3.45 &    10.19 &    1.02 &   0.98 &   $-$0.05 \cr
\noalign{\medskip\hrule\medskip}
}
\endtab

\begfig 5.5 cm
\figure{5}{Response in percent of the 249.7~nm equivalent width (solid line)
and
the population of the ground state at a representative line-forming
depth (dotted line) as the photoionisation cross sections are increased
 by a factor two for each atomic level at a time. Results are
for the HD140283 atmosphere and the ten lowest atomic levels
}
\endfig

\begfig 6.5 cm
\figure{6} {Normalised net rates in the \bi\ continua as functions of depth
in HD140283. Ionisation rates are positive and recombination is negative.
Numbers indicate the lower levels involved. The dotted lines represent the
levels
from number 8 and upwards. Note the different scales in the right and the left
part of the diagram}
\endfig

\begfig 7.5 cm
\figure{7}{The depletion factor of neutral boron ($d_0$) as the different
pumping mechanisms are turned on and off. \hfill\break
{\bf a}: standard atom model (thick line)\hfill\break
 Optical pumping turned off in: \hfill\break
{\bf b}:~All resonance lines (dotted),
{\bf c}:~250~nm, 209~nm, and 183~nm lines (dashed),
{\bf d}:~All resonance lines except 250~nm, 209~nm, and 183~nm (dash-dotted),
{\bf e}:~250~nm (dash-dot-dotted),
{\bf f}:~209~nm (long-dashed)
{\bf g}:~183~nm (full-drawn)\hfill\break
The right panel shows
the same thing when $J=B$ in all photoionisation continua
}
\endfig

\titlea{Results}

Using the methods described, NLTE calculations were made for the
three stellar models. The abundances used were the
LTE results of the existing boron abundance determinations:
2.6 ($\pm 0.3$) for the Sun (Kohl et al. 1977), 1.90 for Procyon (upper limit
of LLE93) and $-0.10$ for HD140283 (upper limit of
DLL92).
The equivalent widths of the \bi\ resonance lines studied
show significant differences
between NLTE and LTE for the HD140283 and
Procyon cases -- the lines are
weaker in NLTE -- but very small differences for the Sun.
To get the appropriate conversions between LTE and NLTE
abundances, calculations were made for a range of abundances, and
the resulting curves of growth were used to interpolate the quantities
in Tables 4-6. These tables allow translation of
 an observed equivalent width $W$ or a deduced
LTE abundance $A_{\rm B}^{\rm LTE}$ to a "true" abundance
$A_{\rm B}$ with a NLTE abundance correction $\Delta A_{\rm B}$.
 Applying the NLTE corrections
on the literature abundance results just cited gives the following
corrected abundances: 2.62 for the Sun, 2.29 for Procyon, and 0.46 for
HD140283

The departures from LTE in the three atmospheric models
are illustrated with departure coefficients for the level populations
and line source functions in the plots
of Fig. 4. As expected from the equivalent-width results,
the NLTE deviations are small in the Sun, and larger
in Procyon and HD140283. For the ultraviolet resonance lines, we see
a line source function much greater than the local Planck function.
Since $S_l/B = b_u/b_l$ in this case, there is a significant overexcitation
-- as is obvious from the plotted $b_i$-coefficients.
There is also overionisation, resulting in a depletion of neutral
boron and thus a decrease of the line opacity. Both these effects
combine to make the resonance lines weaker compared to the LTE case.
(The lines are still in absorption in integrated light. There is a small
 emission
contribution from the limb in Procyon  and HD140283.)

In Fig. 4, the formation region of the 249.7~nm line is indicated
 with ticks
at line-centre total optical depth 0.1 and 1. For HD140283, the contribution
function of Magain (1986) peaks at $\lg \tau_5 = -1.0$.

\titlea{Interpretation of the results}
In the following the nature of the NLTE effects will be investigated
 and the reliability
of the results will be assessed. This is done by a number
of test runs made with different perturbations on the model atom
and atmospheres.
The main subject here will be to understand the results for HD140283.
One method to diagnose the situation is
a "multi-MULTI" run, where each oscillator strength and collisional
cross section in the model atom is doubled in turn and the results
-- line equivalent widths and population numbers --
are compared those of the unperturbed model atom. The procedure
was introduced and described by Carlsson et al. (1992).
The current multi-MULTI run was kindly administered by Dr.~M.~Carlsson
for the HD140283 atmosphere. Some of the relevant results are
presented in Fig. 5. They will be discussed in the following.

\titleb{Line source functions}
A first conclusion is that the resonance doublets of observational interest,
209~nm and 250~nm, have source functions well described by a two-level
atom approximation: $$S_l = (1-\epsilon) \bar J +  \epsilon B$$
with $\epsilon$ being the fraction of de-excitations that are collisional,
 and $\bar J$ is the mean intensity
averaged over the line profile, (e.g. Mihalas 1978, p.337).
This conclusion rests on the facts that the rates in the transitions
connecting to the
upper level of these lines are totally dominated by the radiative rates
in the lines, and that the line source functions are virtually unchanged
in the different experiments described here. The latter point
was evident in the multi-MULTI run: when all other line oscillator strengths
were doubled in turn, the 249.7~nm equivalent width never changed
with more than 1\,\%.
These resonance lines are dominated by scattering with $\epsilon \sim 10^{-5}$.
 This means that the line source functions $S_l$
should be equal to the local mean intensities $J$, something which is
indeed seen in the lower HD140283 panel of Fig. 4. What is plotted
there is in fact $J_C$, the mean intensity in the continuum, which does
not differ much from $\bar J$ in the lines as long as they are weak.
The conclusion
is that $J$ and thus
$S_l$ are set by the background opacities for the lines in question.

\titleb{Cause of the overionisation}
The line opacities are determined by the populations of the lower levels, which
in this case is the ground state. Since the absolute majority of the
neutral boron atoms in the line-forming regions are in the ground state,
the important thing is the ionisation balance. The task then
is to explain the overionisation effect, demonstrated
by the neutral depletion factor $d_0$ in Fig. 4.

A look at the most important net rates in the transitions connecting to the
\bii\ level (Fig. 6)
shows that the net contribution is dominated by ionisation from the
upper excited levels below $\lg \tau_5 = -0.3$. Higher up,
the ionisation net contributions from the lower excited
levels, primarily from levels 4, 5, and 6, become important. This
observation can be used to set up a hypothesis regarding the mechanisms
behind the overionisation. Going outwards through the atmosphere,
overionisation sets in when the  resonance lines with $\lambda < 180$~nm
begin to be optically pumped as their radiation fields deviate from the local
Planckian value. This leads to increased photoionisation from the
higher excited levels,
whose continua have a much richer radiation field than the ground state
(but not necessarily with $J>B$), so that the ionisation balance is shifted
relative to the LTE value.
Higher up in the atmosphere, pumping in the longer-wavelength doublets
becomes more important and the most important net contributions come
from levels 4 and 6, which are directly influenced by pumping, and
level 5, which is coupled to levels 3, 4, and 6.

A series of experiments confirm this picture. In Fig. 7, the $d_0$ depletion
coefficient is shown as different
line pumping mechanisms are turned on and off in various combinations. A
line pump is turned off by treating the transition as a fixed rate
calculated with mean intensities given by the local Planckian
values.
It is obvious how the absence of pumping in the 183~nm, 209~nm and 250~nm
lines decreases
the overionisation above $\lg \tau_5 = -0.5$, with the 209~nm pump being the
most important.
At greater depths, the overionisation
disappears when the pumping in the shorter-wavelength resonance lines
is turned off. The
procedure is repeated in the right panel of Fig. 7, now with the
radiation fields in all the photoionisation continua
forced to local Planckian values. It is interesting to note that, while
the overionisation decreases somewhat,
it is still substantial in this case.

\titleb{Results of multi-MULTI perturbation test}
The multi-MULTI run showed that the equivalent widths of interest were
unsensitive to perturbations in oscillator strengths and bound-bound cross
sections, while changes in some bound-free cross sections had influence
on the line strengths. These results can now be interpreted within the
picture proposed in the preceding sections.
The insensitivity of the 249.7~nm equivalent widths to perturbations
of other $f$-values and bound-bound cross sections is due to the two-level
nature of the line (no
change in $S_l$) plus the fact that the effectiveness of
the optical pumping in the resonance lines is independent of $f$-value as
long as the
line is weak enough (thus no change in line opacity).

Looking at Fig. 5, one sees that an increase in the photoionisation
cross sections of the ground state (0 and 1)
decreases the overionisation. This can be explained
as a consequence of the stronger coupling between the ground
state and the continuum, since the radiation field at the ground
state photoionisation edge does not deviate much from $B$ as compared
with the other levels (cf. Fig. 2). The effective ionisation edge
of level 2 is at even shorter wavelength where the radiation field
is weaker and even closer to Planckian.
 The weak response of level 3 is explained by the
systematically low ionisation cross sections of all the $^2S$ levels.
Levels 4, 5, and 6 seem to be most important for the overionisation in
the lineforming regions, as was indicated by Fig. 6.

\titleb{Unimportance of photon suction}
Carlsson et al. (1992) demonstrated the importance of the high-lying Rydberg
levels for the recombination flow of neutral magnesium. For the
alkali atoms, Bruls et al. (1992) have showed in convincing detail that
this recombination flow, boosted by photon loss in high-probability
infrared lines ("photon suction"), counteracts the overionisation
from the lower levels so effectively that the net result is actually an
 underionisation.
Apparently this effect is not as effective in the current situation.
To see whether such photon suction is of any importance at all, test runs
were made where all radiative transitions between excited levels (all
in the red or infrared) were removed. The resulting changes in the ionisation
are small, as are the corresponding changes in line equivalent widths --
$W$ for the 249.7~nm line {\it decreases} with less than 2\,\% in HD140283.
Apparently the effect of these lines is not to increase the recombination
flow via photon suction but instead to increase the overionisation somewhat
by their effect of coupling the higher levels to each other.

\titleb{What matters most?}
More experiments help to show the relative importance of the different
processes just discussed.
Inspection of Table 7, labels {\bf A} through {\bf F}, shows that it
is mainly the pumping in the observed line together with
the pumping in all other resonance lines that give the
appreciable NLTE effect on the 249.7~nm equivalent width for HD140283.
The non-Planckian radiation field in the photoionisation continua does
not contribute much to the overionisation (compare {\bf A} and {\bf B}).

\titleb{Procyon and the Sun}
Looking at the other stars it seems that the processes in the Procyon
atmosphere are similar to those in HD140283.
 In the Sun, all deviations from LTE are much
smaller. The significant difference seen in the behaviour of the 209~nm and
the 250~nm lines (Fig. 4) is due to the different radiation temperatures at
those
two wavelengths. The 249~nm line is situated just on the short wavelength
side of the photoionisation
edge of Mg\,{\sc i} at 251.2~nm, while the opacities are lower at 209~nm.
In the metal-poor atmosphere, the difference in opacities between
these wavelengths is much smaller. This is, however, partly an artificial
phenomenon due to the omission of background line opacities.
The lower panels of Fig. 4 show how the approximation $S_l = J_C$ fails
as we go from HD140283, over Procyon, to the Sun. This comes mainly from
the increasing line strength -- the lines are reasonably well described
by a two-level source function also in the Sun.

Tables 8 and 9 contains the same experiments regarding the importance
of the different processes as were discussed in the previous section for
HD140283.

\begtabfull
\tabcap{7}{HD140283, equivalent widths and NLTE effects. $d_0$ and $S_l/B$ are
given at $\lg\tau_5=-1.1$. See text for
further explanations. \hfill\break
{\bf A~}Standard parameters \hfill\break
{\bf B~}Photoionising radiation fields set to local Planckian \hfill\break
{\bf C~}Collisional coupling between ground state and continuum increased
to enforce Saha equilibrium \hfill\break
{\bf D~}Collisional rates in 250~nm doublet increased to enforce
Boltzmann excitation balance there \hfill\break
{\bf E~}Collisional b-b rates set to large values to assure
Boltzmann excitation balance throughout atom \hfill\break
{\bf F~}B + D \hfill\break
{\bf G~}Photoionising radiation field increased ten times\hfill\break
{\bf H~}Photoionising radiation fields in ultraviolet increased
following Bikmaev \& Steenbock (1993) \hfill\break
{\bf I~}Schematic line blanketing included in background
opacities of resonance lines \hfill\break
{\bf J~}Collisional ionisation rates
increased with a factor 100 \hfill\break
{\bf K~}No electron collisional ionisations \hfill\break
{\bf L~}Collisional cross sections among lower levels increased with a factor 5
to fit experimental data \hfill\break
{\bf M~}All collisional cross sections according to the Van Regemorter (1962)
approximation \hfill\break
{\bf N~}Collisions with neutral hydrogen including according to Drawin (1969)
and Steenbock \& Holweger (1984)\hfill\break
}
\halign{#\hfil\quad\quad &#\hfill\quad\quad\ &\hfill#\ &\hfill#\
&\hfill#\quad\quad\ &\hfill#\ &\hfill#\ &\hfill#\ &\hfill#\ \cr
\noalign{\hrule\medskip}
\noalign{~~~~~~~~~~~~~~~~~~~209.0~nm~~~~~~~~~~~~~~~249.7~nm }
         & $d_0$ & $W$      & $W^*\over W$  & $S_l\over B$ & $W$ & $W^*\over W$
 & $S_l\over B$    \cr         &       & [pm]     &               &
              & [pm] \cr
\noalign{\medskip\hrule\medskip}
{\bf A}  &     1.7 &    0.16 &   4.1 &  11.2 &    0.23 &   3.4 &   6.0 \cr
{\bf B}  &     1.6 &    0.17 &   3.8 &  11.2 &    0.25 &   3.2 &   6.0 \cr
{\bf C}  &     1.0 &    0.29 &   2.3 &  11.2 &    0.41 &   1.9 &   6.1 \cr
{\bf D}  &     1.7 &    0.17 &   3.9 &  11.2 &    0.46 &   1.7 &   1.0 \cr
{\bf E}  &     1.0 &    0.64 &   1.0 &   1.0 &    0.77 &   1.0 &   1.0 \cr
{\bf F}  &     1.0 &    0.65 &   1.0 &   1.0 &    0.79 &   1.0 &   1.0 \cr
{\bf G}  &     4.9 &    0.08 &   8.1 &  11.1 &    0.11 &   7.2 &   6.0 \cr
{\bf H}  &     2.2 &    0.13 &   4.9 &  11.2 &    0.19 &   4.2 &   6.0 \cr
{\bf I}  &     1.7 &    0.18 &   4.1 &  10.7 &    0.28 &   2.9 &   5.1 \cr
{\bf J}  &     2.2 &    0.13 &   5.0 &  11.2 &    0.19 &   4.2 &   6.0 \cr
{\bf K}  &     1.6 &    0.18 &   3.6 &  11.2 &    0.26 &   3.1 &   6.0 \cr
{\bf L}  &     1.7 &    0.16 &   4.1 &  11.2 &    0.23 &   3.4 &   6.0 \cr
{\bf M}  &     1.7 &    0.16 &   4.1 &  11.2 &    0.23 &   3.4 &   6.0 \cr
{\bf N}  &     2.4 &    0.15 &   4.3 &   9.5 &    0.18 &   4.4 &   6.1 \cr
\noalign{\medskip\hrule\medskip}
}
\endtab

\begtabfull
\tabcap{8}{Procyon, equivalent widths and NLTE effects.  $d_0$ and $S_l/B$ are
given at $\lg\tau_5=-1.3$. See Table 7 for
explanations \hfill\break
}
\halign{#\hfil\quad\quad &#\hfill\quad\quad\ &\hfill#\ &\hfill#\
&\hfill#\quad\quad\ &\hfill#\ &\hfill#\ &\hfill#\ &\hfill#\ \cr
\noalign{\hrule\medskip}
\noalign{~~~~~~~~~~~~~~~~~~~209.0~nm~~~~~~~~~~~~~~~249.7~nm }
         & $d_0$ & $W$      & $W^*\over W$  & $S_l\over B$ & $W$ &
$W^*\over W$ & $S_l\over B$    \cr
         &       & [pm]     &               &              & [pm] \cr
\noalign{\medskip\hrule\medskip}
{\bf A}  &     1.5 &    1.15 &   2.3 &   6.0 &    1.28 &   2.1 &   2.3 \cr
{\bf B}  &     1.5 &    1.17 &   2.3 &   6.0 &    1.30 &   2.1 &   2.3 \cr
{\bf C}  &     1.0 &    1.74 &   1.5 &   5.9 &    1.95 &   1.4 &   2.2 \cr
{\bf D}  &     1.5 &    1.18 &   2.2 &   5.9 &    1.85 &   1.5 &   1.0 \cr
{\bf E}  &     1.0 &    2.64 &   1.0 &   1.0 &    2.73 &   1.0 &   1.0 \cr
{\bf F}  &     1.0 &    2.65 &   1.0 &   1.0 &    2.73 &   1.0 &   1.0 \cr
{\bf G}  &     2.6 &    0.83 &   3.2 &   5.7 &    0.86 &   3.2 &   2.2 \cr
{\bf I}  &     1.3 &    1.54 &   1.7 &   4.5 &    1.77 &   1.5 &   1.7 \cr
{\bf J}  &     2.4 &    0.90 &   2.9 &   5.9 &    0.97 &   2.8 &   2.2 \cr
{\bf K}  &     1.3 &    1.39 &   1.9 &   5.9 &    1.52 &   1.8 &   2.2 \cr
{\bf L}  &     1.5 &    1.16 &   2.3 &   5.9 &    1.28 &   2.1 &   2.3 \cr
{\bf M}  &     1.5 &    1.15 &   2.3 &   6.0 &    1.28 &   2.1 &   2.3 \cr
{\bf N}  &     2.4 &    0.97 &   2.7 &   5.1 &    0.96 &   2.9 &   2.3 \cr
\noalign{\medskip\hrule\medskip}
}
\endtab

\begtabfull
\tabcap{9}{Sun, equivalent widths and NLTE effects. $d_0$ and $S_l/B$ are
given at $\lg\tau_5=-2.8$. See Table 8 for
explanations \hfill\break
}
\halign{#\hfil\quad\quad &#\hfill\quad\quad\ &\hfill#\ &\hfill#\
&\hfill#\quad\quad\ &\hfill#\ &\hfill#\ &\hfill#\ &\hfill#\ \cr
\noalign{\hrule\medskip}
\noalign{~~~~~~~~~~~~~~~~~~~209.0~nm~~~~~~~~~~~~~~~249.7~nm }
         & $d_0$ & $W$      & $W^*\over W$  & $S_l\over B$ & $W$ &
$W^*\over W$ & $S_l\over B$    \cr
         &       & [pm]     &               &              & [pm] \cr
\noalign{\medskip\hrule\medskip}
{\bf A}  &     1.0 &    7.34 &   1.1 &   2.5 &    6.74 &   1.0 &   0.8 \cr
{\bf B}  &     1.0 &    7.34 &   1.1 &   2.5 &    6.75 &   1.0 &   0.8 \cr
{\bf C}  &     1.0 &    7.41 &   1.1 &   2.5 &    6.79 &   1.0 &   0.8 \cr
{\bf D}  &     1.0 &    7.37 &   1.1 &   2.5 &    6.74 &   1.0 &   1.0 \cr
{\bf E}  &     1.0 &    7.84 &   1.0 &   1.0 &    6.79 &   1.0 &   1.0 \cr
{\bf F}  &     1.0 &    7.84 &   1.0 &   1.0 &    6.79 &   1.0 &   1.0 \cr
{\bf G}  &     1.3 &    7.16 &   1.1 &   2.4 &    6.44 &   1.1 &   0.8 \cr
{\bf I}  &     1.0 &    7.61 &   1.0 &   1.7 &    6.33 &   1.1 &   1.2 \cr
{\bf J}  &     1.0 &    7.31 &   1.1 &   2.5 &    6.72 &   1.0 &   0.8 \cr
{\bf K}  &     1.0 &    7.36 &   1.1 &   2.5 &    6.75 &   1.0 &   0.8 \cr
{\bf L}  &     1.0 &    7.35 &   1.1 &   2.5 &    6.75 &   1.0 &   0.8 \cr
{\bf M}  &     1.0 &    7.33 &   1.1 &   2.5 &    6.74 &   1.0 &   0.8 \cr
{\bf N}  &     1.0 &    7.56 &   1.0 &   1.5 &    6.62 &   1.0 &   0.9 \cr
\noalign{\medskip\hrule\medskip}
}
\endtab

\begtabfull
\tabcap{10}{Change of the NLTE effect in the resonance lines as the
atmospheric parameters are perturbed. The change is expressed
as $\lg W_0/W_0^* - \lg W_1/W_1^*$, where indices 0 and 1 refer
to the standard and the perturbed model, respectively. The
numbers thus give the approximate change in dex of the NLTE correction on
the abundances
}
\halign{#\hfil\quad &\hfill#\quad &\hfill#\quad &\hfill#\quad \cr
\noalign{\hrule\medskip}
        & $\Delta$\teff &$\Delta \lg g$ & $\Delta$\feh \cr
\noalign{\medskip\hrule\medskip}
HD140283\cr
         & $+$100~K & $+$0.4 & $+$0.4 \cr
\noalign{\medskip}
209.7~nm &   $+$0.07  & $-$0.12 &  $-$0.02 \cr
249.7~nm &   $+$0.07  & $-$0.09 &  $-$0.02 \cr
\noalign{\medskip\hrule\medskip}
Procyon\cr
         & $+$100~K & $+$0.4 & $-$0.4 \cr
\noalign{\medskip}
209.7~nm &   $+$0.03 &  $+$0.00 &  $+$0.09 \cr
249.7~nm &   $+$0.03 &  $+$0.01 &  $+$0.10 \cr
\noalign{\medskip\hrule\medskip}
}
\endtab

\titlea{Reliability of the results}
How sensitive are the NLTE effects found in the preceding section
on the input data and the assumptions made?
Apart from the multi-MULTI run,
 the reliability of the
results was tested by repeating the calculations with different
perturbations of the input fluxes, atomic data, etc.  The results for most
of these tests, in the form
of equivalent widths ($W$), ratios between LTE and
NLTE equivalent widths ($W^*W$), and neutral depletion
coefficients $d_0$ plus $S_l/B$ values at typical line-forming depths
are presented in Tables
7-9 under labels {\bf A} -- {\bf N}. These letters
will be used for reference in the following. The standard result is
given under the label {\bf A}.

\titleb{Radiation fluxes}

The NLTE behaviour of neutral boron is sensitive
to the photoionising fluxes and to the background fluxes in the
resonance lines.
For the photoionisation fluxes, one may get some confidence
that the model atmosphere results reproduce the observed fluxes well when
smoothed to about 5~nm from
Edvardsson et al. (1993) (Sun and Procyon) and LLE93 (Procyon).
For HD140283, Bikmaev \& Steenbock (1993) used IUE data to
deduce an ultraviolet excess between 175~nm and 320~nm
compared to what the model atmospheres give. This excess
grows to $+$0.53 dex at the shortest wavelength.
To assess the sensitivity of the results on these fluxes,
calculations were made where all mean intensities in
the photoionisation calculations were increased with a
factor of 10 in the wavelength range where the
atmospheric models gave $J/B > 1$, label {\bf G} in Tables 7-9.
For HD140283 the
Bikmaev \& Steenbock (1993) data was used in the same way, results
in Table 7, label {\bf H}.
These corrections were applied as constant factors to the
radiation field at all depth points. The result is a
decrease in line strength as the overionisation is boosted.

The background fluxes in the resonance lines are a potential
source of errors due to the
omission of lines in the background opacities. Line blanketing will
decrease the strength of the optical pumping in the ultraviolet
resonance lines. In order to estimate the
importance of this to some extent, calculations were made where
the resonance
lines were treated as transitions with fixed rates for which the same
mean intensities as for the photoionisations were used.
All resonance lines were treated this way except the
249.68~nm and 208.96~nm lines. These lines are effectively
unblended according to the observed spectrum of DLL92 and
the line list of Johansson et al. (1992).
The procedure gives a reasonable representation for the line blanketing
effect smoothed over a wavelength interval,
but there is no guarantee that the line blanketing at
the wavelength of each specific line is representative
for the regional mean.
The resulting reduction of the NLTE effects, as presented under
label {\bf I},
is not so great as to affect the general conclusions regarding HD140283,
though Procyon is more influenced. The effect of line blanketing is
to decrease the overionisation via decreased pumping in the resonance
lines and also to decrease the line source functions of the unblanketed
components by their coupling to their doublet partners.
To further improve the situation, a detailed line list for the
spectral regions around the
resonance lines must be included in the NLTE calculations.

The mean flux around 249~nm produced by the model atmospheres program
is consistent with IUE observations of Procyon according to LLE93.
At 209~nm, the cited study of Bikmaev \& Steenbock (1993) resulted in a
correction
factor of two for the mean flux of HD140283. Such an excess could
be due to NLTE effects (or other errors)
in the continuous opacities or the effects of granulation, and could of
course further increase the deviations from LTE in the \bi\ lines.
It is not, however, straightforward to assess the impact of background
opacity changes,
since lower continuous opacities to first order lead to stronger absorption
lines, while the increased ultraviolet flux at the same times should
make these lines weaker due to pumping and overionisation. In HD140283,
the LTE continuous opacities (as computed by the code) at 250~nm and 209~nm
 are almost exclusively due to hydrogen in different forms.

Finally it should be noted that the photoionisation cross sections
of several of the excited levels show a strong resonance peak
around 135~nm. There is a possibility that chromospheric emission
in O\,{\sc i} 136~nm or C\,{\sc ii} 134~nm could influence the
ionisation equilibrium through this resonance.

\titleb{Model atom details}
Some tests of the sensitivity on the uncertainties of the model atom
were also made.

An increase of the electron collisional ionisation
rates with a factor of 100 increases the NLTE effects in HD140283
and Procyon ({\bf J}). This is probably due to the fact that
the coupling of the continuum to the excited levels that are affected
by the optical pumping is increased. For the same reason, the overionisation
decreases when this collisional coupling is removed ({\bf K}). The electron
collisional ionisation is then
a possibly significant factor of uncertainty, since the cross-section values
used must be considered as very approximate.

As mentioned earlier, the importance of bound-bound inelastic collisions with
 hydrogen atoms for problems such as the current one is
a matter of debate. One may think that the omission here of
such collisional processes may lead to an overestimation of
the NLTE effect. However, tests where such collisions were included,
 using the rate coefficients of Drawin (1969) in the
Steenbock \& Holweger (1984) formulation, actually showed an increased
overionisation and still smaller resonance line strengths ({\bf N}).
A closer look at the results, and a few more experiments, made it probable
that this is due to the increased coupling between the $^2S$ and
the $^2P$ levels as well as between the $^2P$ and the $^2D$ levels.
This increases the overionisation since the $^2S$ levels have so
low photoionisation cross sections and the $^2D$ levels are otherwise
unaffected by the resonance line pumping.

Turning to the bound-bound electron collisions,
 we are happy to have precision OP cross sections
for the transitions between the lower levels which directly couple
to the observed lines.
Remembering the possible discrepancy between these data and experiments
(see section 3.6), the cross sections were increased with a factor of
5 to better fit the experiments mentioned.
The results, label {\bf L}, are effectively unchanged.

An experiment was also done where {\it all} collisional rates were calculated
according to the Van Regemorter (1962) approximation as described in
section 3.6. This could
possibly give a more correct balance between the different collisional
rates than mixing the two sources. The results ({\bf M}) do not, however,
show any difference.

Apparently, uncertainties in the collisional cross sections can be of
importance for transitions between the higher excited levels, while
the results are not sensitive to the uncertainty in the
collisional rates between the lower levels.

The lone bound level in the quartet system is worth a special look. This
metastable level cannot be ionised directly to the ground state of
\bii, the effective ionisation edge is at 135~nm instead of at
260~nm. If there was an ionisation channel, perhaps via autoionising levels,
that allowed ionisation by photons around 260~nm wavelength (which
is close to the maximum in $J/B$ in Fig. 2), this could
significantly increase the overionisation. This potential effect was discovered
during earlier phases in this work when there was no photoionisation
cross section available for this level and various substitutes were tested.

As a final check of the results, large collisional rates among the bound levels
were combined
with the local Planckian photoionisation radiation fields to see
if the LTE value was regained. As seen in Tables 7-9, label {\bf H},
this indeed happened.

In conclusion, the most important uncertainties in the atomic model seem to be
the possibility
of photoionisation from level 2 via alternative channels, the collisional
ionisation from the higher levels and the bound-bound
collisional coupling between the higher levels (in this order).

\titleb{Atmospheric parameters}
The sensitivity of the NLTE effect on the basic atmospheric parameters was
investigated using model atmospheres with perturbed values
of the effective temperatures, surface gravities and metallicities.
Table 10 shows the logarithmic changes in the $W/W^*$ values.
Uncertainties in the fundamental parameters are of course of importance
for the analysis, but they cannot remove
all of the NLTE effects for Procyon and HD140283.

\titleb{Summary}
The uncertainties in the model atom or in the background fluxes may well change
the current results regarding the size of the NLTE effect. However, it
seems unlikely that this could altogether remove the significant
departures from NLTE for the resonance lines of neutral boron in stars
more metal-poor and/or hotter than the Sun.
An argument for this is the demonstration that both of the basic processes
responsible for the departures from LTE, pumping in all resonance lines and
pumping in the observed lines, produces a significant departure alone.

The effect of proper inclusion of line blanketing will work in
the direction of decreasing the departures from LTE. There is a hint
that improving the UV fluxes will work the other way, while it is difficult
to assess which way improved atomic data will change the results.
The nominal result for HD140283 was that $W^*/W = 3.4$ for the 249.7~nm
line. After considering the experiments of Table 7, I estimate a minimum value
for that quantity of ${W^*/W} = 2.9$ (experiment {\bf I}),
and a maximum value of $W^*/W = 4.2$ (experiment
{\bf H}). These values transcribe to approximately similar factors in
abundance.
Thus it seems reasonable to state that the NLTE abundance
correction for HD140283,
given the current atmospheric model, should lie between
$+0.46$ and $+0.62$ dex.

Far more serious errors of analysis than discussed so far may be
introduced by photospheric
inhomogeneities, granulation. DLL92 propose that the agreement
between predicted
and observed fluxes at 249~nm for HD140283 indicates that the inhomogeneities
do not have a large effect. Studying fluxes at shorter wavelengths,
Bikmaev \& Steenbock (1993) came to the conclusion that there is a discrepancy.
The impact of granulation on the line formation processes must
be carefully tested with three-dimensional simulations and
line-calculations before any conclusions concerning the effects
on boron abundance analysis are possible. The prediction of a
substantial NLTE effect in plane-parallel atmospheres could also indicate
that granulation may, however, be a very important factor for the
formation of these
lines -- more important than for continuum formation.
After all, NLTE effects are non-locality effects, and spectral lines
vulnerable to these should also be sensitive to the physical conditions
in the horizontal direction!

\titlea{Conclusions}
The departures from LTE found in this study for the metal-poor star
HD140283 and for Procyon can be described as an optical pumping
effect in the resonance lines of observational interest and a general
depletion of neutral boron. The overionisation is driven by optical
pumping in all the ultraviolet resonance lines with the 209~nm doublet
being the most important.
It is interesting that there is overionisation even with $J=B$ in
the photoionisation continua. It is also notable that some of the experiments
with overall increased collisional rates actually increased the departures
from LTE -- contrary to the standard notion that collisional processes
drive populations towards Boltzmann and Saha distributions.

The predicted NLTE effect in the Sun is far smaller than other uncertainties
caused by continuum definition, blending, etc. (Kohl et al. (1978) give
their uncertainty in the abundance as $\pm 0.3$~dex.) There is no reason to
change the standard solar value for the boron abundance due to NLTE
effects.

If the NLTE abundance correction for Procyon is applied
to the LLE93 result, the boron depletion of this star relative to the
Sun is decreased, possibly even removed. This could put stronger
constraints on the mechanisms behind the lithium and beryllium
depletion observed in Procyon, cf. LLE93.

Applying the abundance correction for HD140283 on the DLL92 result
for boron raises the B/Be ratio for this star to 34.
This is well above the minimum requirement for spallation production
of beryllium: ${\rm B}/{\rm Be} = 10$ from DLL92. It is even close
to the maximum value allowed for spallation production of boron and beryllium
according to Fig. 5 these authors.
The uncertainties in radiation fluxes and atomic quantities, as discussed
above, may change the exact figure, but could hardly change the general
conclusion. This may be adjusted if the DLL92 result really is
an upper limit due to an unidentified blend (not at all likely according
to LLE93), or if beryllium (but not boron) has been strongly depleted in the
HD140283 atmosphere.

The measured beryllium abundance is probably not as important as a factor
of uncertainty
for the B/Be ratio, since it is derived from lines of Be\,{\sc ii}. The
LTE value of $n_{\rm Be^+}/n_{\rm Be}$ in the HD140283 atmosphere is about
 60\,\%,
according to calculations kindly provided by Olle Morell.
An overionisation effect in HD140283 could then, in the very extreme case,
increase the number of Be ions by two thirds.
Any NLTE effect in the excitational
balance of the Be lines would probably not be greater than that
found here in boron. The
observed boron resonance lines are situated in the spectral region where
$J/B$ reaches its general maximum -- we expect less deviations
of the mean intensity from the local Planckian value
for the Be\,{\sc ii} resonance lines at 313~nm (cf. Fig. 2).
Note that the possible effects of
overionisation and of non-Planckian source functions in these
ultraviolet lines would
seem to work in opposite directions on the resulting absorption line strengths,
rather than being synergetic like in neutral boron. An overionisation
strengthens the lines, leading to an overestimation of abundance, while
an overexcitation leads to an underestimation of abundance in an LTE analysis.

This NLTE study of boron thus adds weight to
the interpretation of beryllium in metal-poor stars as being
produced by spallation processes and not being of cosmological origin.

\acknow{ I am indebted to Mats Carlsson for very valuable discussions,
help with MULTI and IDL routines, and for the multi-MULTI experiment.
Keith Berrington is thanked for assistance with atomic data, as is Sultana
Nahar.
The kind help of Bengt Edvardsson (model atmospheres),
and Olle Morell (Be ionisation balance) is acknowledged.
Bengt Gustafsson proposed this study and
gave valuable criticism and comments on the manuscript.
The paper benefited from discussions
with Kjell Eriksson, G\"oran Hammarb\"ack, and Andreas Redfors.
The critical comments of the referee, Robert J. Rutten, improved this paper
considerably. This work was
supported by the Swedish Natural Science Research Council. }

\begref{References}

\ref Achmad L., de Jager C., Nieuwenhuijzen H., 1991, A\&A, 250, 445
\ref Allen C.W., 1973, Astrophysical Quantities, The Athlone Press,
University of London
\ref Bashkin S., Stoner Jr. J. O., 1978, Atomic Energy-level and
Grotrian Diagrams
 Volume I. Hydrogen I - Phosphorus XV Addenda, North-Holland Publishing
Company,
Amsterdam
\ref Berrington K.A., Burke P.G., Butler K., et al., 1987,
J. Phys. B: At. Mol. Phys., 20, 6379
\ref Bikmaev I.F., Steenbock W., 1993, in preparation
\ref Bruls J.H.M.J., Rutten R.J., Shchukina N.G., 1992, A\&A, 265, 237
\ref Caccin,  B., Gomez M.T., Severino, G., 1993, AA, 276, 219
\ref Carlsson M., 1986, A computer program for solving multi-level
non-LTE radiative transfer problems in moving or static atmospheres,
Uppsala Astronomical Observatory Report No. 33
\ref Carlsson M., 1991, Global and local methods for 1-D problems;
implementation aspects and CPU-time and memory scalings.
In: Crivellari L., Hubeny I., Hummer D.G. (eds.)
Stellar Atmospheres: Beyond Classical Models. Kluwer Academic Publishers, p. 39
\ref Carlsson M., Rutten R.J., Shchukina N.G., 1992, A\&A, 253, 567
\ref Cunto, W., Mendoza, C., Ochsenbein, F., Zeippen, C.J.: 1993,  A\&A, 275,
L5
\ref Drawin H.-W., 1969, Z. Phys., 225, 483
\ref Duncan D.K., Lambert D.L., Lemke M., 1992, ApJ, 401, 584, (DLL92)
\ref Edvardsson B., Andersen J., Gustafsson B., et al., 1993, A\&A, 275, 101
\ref Gilmore G., Edvardsson B., Nissen P.-E., 1991, ApJ, 378, 17
\ref Gustafsson B., Bell R.A., Eriksson K., Nordlund \AA , 1975, A\&A, 42, 407
\ref Johansson S.G., Litz\'en U., Kasten J., Kock M., 1993, ApJ, 403, L25
\ref Kiselman D., 1992, Reliability of classical abundance analysis:
boron and oxygen.
In: Alloin D., Stasinska G. (eds.) The feedback of chemical evolution on
 the stellar content
of galaxies. L'Observatoire de Paris, p. 35
\ref Kiselman D., 1993a, Implications of departures from LTE and homogeneity
 in the Sun and
solar-type stars, thesis, Uppsala University
\ref Kiselman D., 1993b, A\&A, 275, 269
\ref Kurucz R.L., 1992, Rev. Mex. Astron. Astrofis., 23, 181
\ref Kohl J.L., Parkinson W.H., Withbroe G.L., 1977, ApJ, 212, L101
\ref Kunc J.A., Soon W.H., 1991, J. Chem. Phys., 95, 5738
\ref Lambert D.L., 1993, Physica Scripta, T47, 186
\ref Lemke M., Lambert D.L., Edvardsson B., 1993, PASP, 105, 468, (LLE93)
\ref Magain, P., 1986,  A\&A, 163, 135
\ref Mihalas, D., Stellar atmospheres, W.H. Freeman and Company, San Francisco
\ref Nakazaki S., Berrington K.A., 1991, J. Phys. B: At. Mol. Phys., 24, 4263
\ref Pagel B.E.J., 1991, Nat, 354, 267
\ref Rutten R.J., 1990, Sun-as-a-star line formation. In: Wallerstein G. (ed.)
ASP Conference Series, Sixth Cambridge Workshop on Cool Stars, Stellar Systems,
and the Sun. Astronomical Society of the Pacific, p. 91
\ref Scharmer G.B., 1981, ApJ, 249, 720
\ref Scharmer G.B., Carlsson M., 1985, J. Comp. Phys., 59, 56
\ref Seaton M.J., 1987, J. Phys. B: At. Mol. Phys., 20, 6363
\ref Steenbock W., Holweger H., 1984, A\&A, 130, 319
\ref Uns\"old A., 1955, Physik der Sternatmosf\"aren, Springer-Verlag,
Berlin, G\"ottingen, Heidelberg
\ref Van Regemorter H., 1962, ApJ, 136, 906
\endref
\bye